\documentclass[pra,aps,amsmath,amssymb,twocolumn,showpacs]{revtex4}
\usepackage{graphicx}

\begin{document}
\title{Giant acceleration
in slow-fast space-periodic Hamiltonian systems}
\author{D.V. Makarov, M.Yu.~Uleysky}
\affiliation{Laboratory of Nonlinear Dynamical Systems,\\
V.I. Il'ichev Pacific Oceanological Institute of the Russian Academy of
Sciences,\\
690041 Vladivostok, Russia}
\date{\today}
\begin{abstract}
Motion of an ensemble of particles in a space-periodic potential well 
with a weak wave-like perturbation imposed is considered. 
We found that slow oscillations of wavenumber of the perturbation
lead to occurrence of directed particle current.
This current is amplifying with time due to giant
acceleration of some particles.
It is shown that giant acceleration is linked with 
the existence of resonant channels 
in phase space.
\end{abstract}
\maketitle

In the recent years considerable interest was devoted to the ratchet effect,
generation of directed particle current in the absence
of any biased forces.
This phenomenon is relevant for wide range of applications,
including controlled photocurrents in semiconductors
\cite{Alekseev,Stevens},
motion of cold atoms in optical lattices \cite{Gommers,Sc,Argonov},
transport of passive tracers
in meandering jet flows in the ocean \cite{Bud},
biological and chemical systems
(see \cite{Reimann} for a comprehensive review).

The ratchet effect in space-periodic Hamiltonian systems is associated with 
the asymmetry of chaotic region
in phase space \cite{FlachPRL,DenFlach,Ketzmerick}.
Recently it was reported about new type of Hamiltonian ratchets,
in which a periodic potential is subjected to a sum of external forces which are periodic
in time and in coordinate \cite{JETPL}.
Each of the forces induces strong but local chaotic diffusion in certain areas of phase space.
This effect is achieved by means of resonance between temporal and spatial
oscillations of the perturbation imposed. 
This resonance
is asymmetric in momentum space,
that provides asymmetry of crossing the separatrix and occurrence
of directed transport.
Combining such forces, we can make finite motion unstable for all range of the particle energy.
By this way, even a weak perturbation of the potential
can activate ballistic current of particles with the lowest initial energies.
A similar effect was used in \cite{Katsouleas} in order
to produce surfatron acceleration.

In the present Letter we demonstrate the mechanism providing
simultaneously generation and {\it giant acceleration}
of directed current by means of a weak external 
perturbation. Possibility of giant acceleration
arises due to adiabatic variations of the perturbation.

Consider an ensemble of non-interacting unit-mass point particles
driven by a wave-like external force.
Motion of each particle is described by the Hamiltonian
\begin{equation}
H=\frac{p^2}{2}-\cos{x}+\varepsilon\cos(\tilde kx+\nu t),
\label{ham0}
\end{equation}
where $\varepsilon\ll 1$, and wavenumber of the perturbation
$\tilde k$ is a slowly-varying parameter
\begin{equation}
\tilde k=k(1+a\cos\Omega t),\quad |a|<1,\quad\Omega\ll 1.
\label{kt}
\end{equation}
Physically this condition corresponds to slow libration
of the external force with respect to axis $x$.
Particle trajectories obey the Hamiltonian equations
\begin{equation}
\dot x=p,\quad \dot p=-\sin{x}+\varepsilon k\sin\phi,
\label{sys}
\end{equation}
where we denote the perturbation phase $\tilde kx+\nu t$ as $\phi$.
Hereafter we shall consider the case when the parameters $k$ and $\nu$ have
sufficiently large positive values, so that the inequality 
$d\phi/dt\gg1$
is satisfied along a particle trajectory, except for some small 
resonant region, where 
\begin{equation}
\frac{d\phi}{dt}=\tilde kp-ka\Omega x\sin{\Omega t}+\nu\simeq0.
\label{res1}
\end{equation}
Outside this region particle dynamics is close to integrable and
can be described using the averaging technique \cite{LanLif1}.
According to (\ref{res1}), the resonant area in phase space
is located along the line given by the following equation
\begin{equation}
p_\text{res}=-\frac{\nu}{k(1+a\cos\Psi)}+
\frac{a\Omega\sin\Psi}{1+a\cos\Psi}x,
\label{str}
\end{equation}
where $\Psi=\Omega t$.
In order to describe the motion inside the resonant region
we derive, using (\ref{sys}) and (\ref{res1}),
the ``pendulum-like'' equation for $\phi$ 
\begin{equation}
\ddot\phi-\varepsilon\tilde k^2\sin\phi+f(x,p,\Psi)=0,
\label{pend-like}
\end{equation}
where $f(x,p,\Psi)$ is treated as a slowly-varying parameter
given by the following equation
\begin{equation}
f(x,p,t)=ak\Omega(2p\sin\Psi+\Omega x\cos\Psi)+\tilde k\sin x.
\label{f}
\end{equation}
Equation (\ref{pend-like}) corresponds to the Hamiltonian of the following form:
\begin{equation}
\tilde H(\dot\phi,\,\phi)=\frac{1}{2}\dot\phi^2+\phi f(x,p,\Psi)+\varepsilon\tilde k^2\cos\phi.
\label{ht}
\end{equation}
If the inequality
\begin{equation}
|f(x,p,\Psi)|<\varepsilon\tilde k^2
\label{crit}
\end{equation}
is satisfied, then the phase portrait 
corresponding to the Hamiltonian (\ref{ht}) contains an oscillating
resonant region in $\phi$--$\dot\phi$ space, bounded by a separatrix.
Falling into resonance (\ref{res1}) corresponds to crossing the separatrix
and entering this region. 
Probability of entering depends on the area of the resonant region
and increases with increasing the difference between the
left- and right-hand sides of (\ref{crit}).
A particle spends any time inside the resonant zone and then
leaves it with strongly increased or decreased energy.
Following \cite{Itin,Neishtadt,Mezic}, we derive
an approximate formula for the energy jump
\begin{equation}
\Delta E=-\varepsilon kp^*\int\limits_{-\infty}^{\phi_*}
\frac{\sin\phi\,d\phi}{\sqrt{2(\tilde H-\phi f^*-\varepsilon k^2\cos\phi)}},
\label{jump}
\end{equation}
where $0<\phi<2\pi$,
$f^*$, $p^*$ and $\phi^*$ are values of $f$, $p$ and $\phi$, respectively,
when the trajectory hits the resonant region.
$\Delta E$ depends extremely on $\phi^*$, therefore
multiple recurrences to the resonant area 
cause chaotic diffusion in phase space \cite{MUM,CNS,Acoust}.

\begin{figure}[!htb]
\begin{center}
\includegraphics[width=0.4\textwidth]{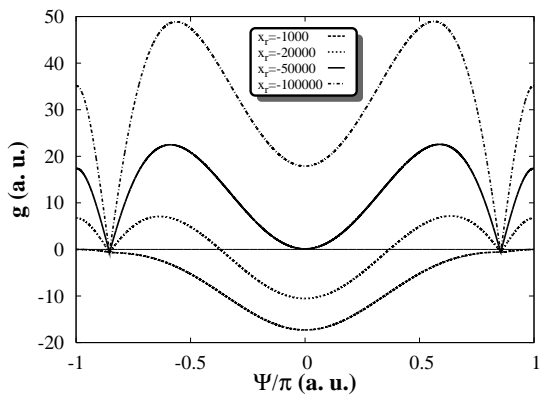}
\caption{Function $g(x,\,\Psi)$ with $x=-1000$,
$x=-20000$, $x=-50000$, and $x=-100000$.}
\label{fig1}
\end{center}
\end{figure}

With fixed values of $k$ and $\nu$, 
the resonant condition (\ref{res1})
has the simplest form
\begin{equation}
kp+\nu\simeq 0.
\label{res2}
\end{equation}
It should be emphasized, that inequality (\ref{crit})
is fulfilled only if a particle is not
far from an extremum of the unperturbed potential.
Hence the condition (\ref{res2})
should be replaced by the following one \cite{CNS}:
\begin{equation}
p(E_\text{res},\,\,x=\pi l)=p_\text{res}\simeq-\frac{\nu}{k},
\label{res21}
\end{equation}
where $l$ is integer. Using (\ref{res21}) we find
the resonant values of energy
\begin{equation}
E_\text{res}=\frac{\nu^2}{2k^2}\mp1.
\label{er}
\end{equation}
Equation (\ref{er}) determines
locations of the chaotic layers in energy space \cite{JETPL,MUM,CNS}.
If the chaotic layer induced by resonance (\ref{res21})
coalesces with the near-separatrix chaotic layer, then
the chaotic sea formed has much larger width
in the lower half-plane of phase space than in the upper one \cite{JETPL}.
It follows from the asymmetry of the condition (\ref{res21}) in momentum space
and implies the prevalence of
chaos-induced particle flights towards $x=-\infty$.

\begin{figure}[!t]
\begin{center}
\includegraphics[width=0.44\textwidth]{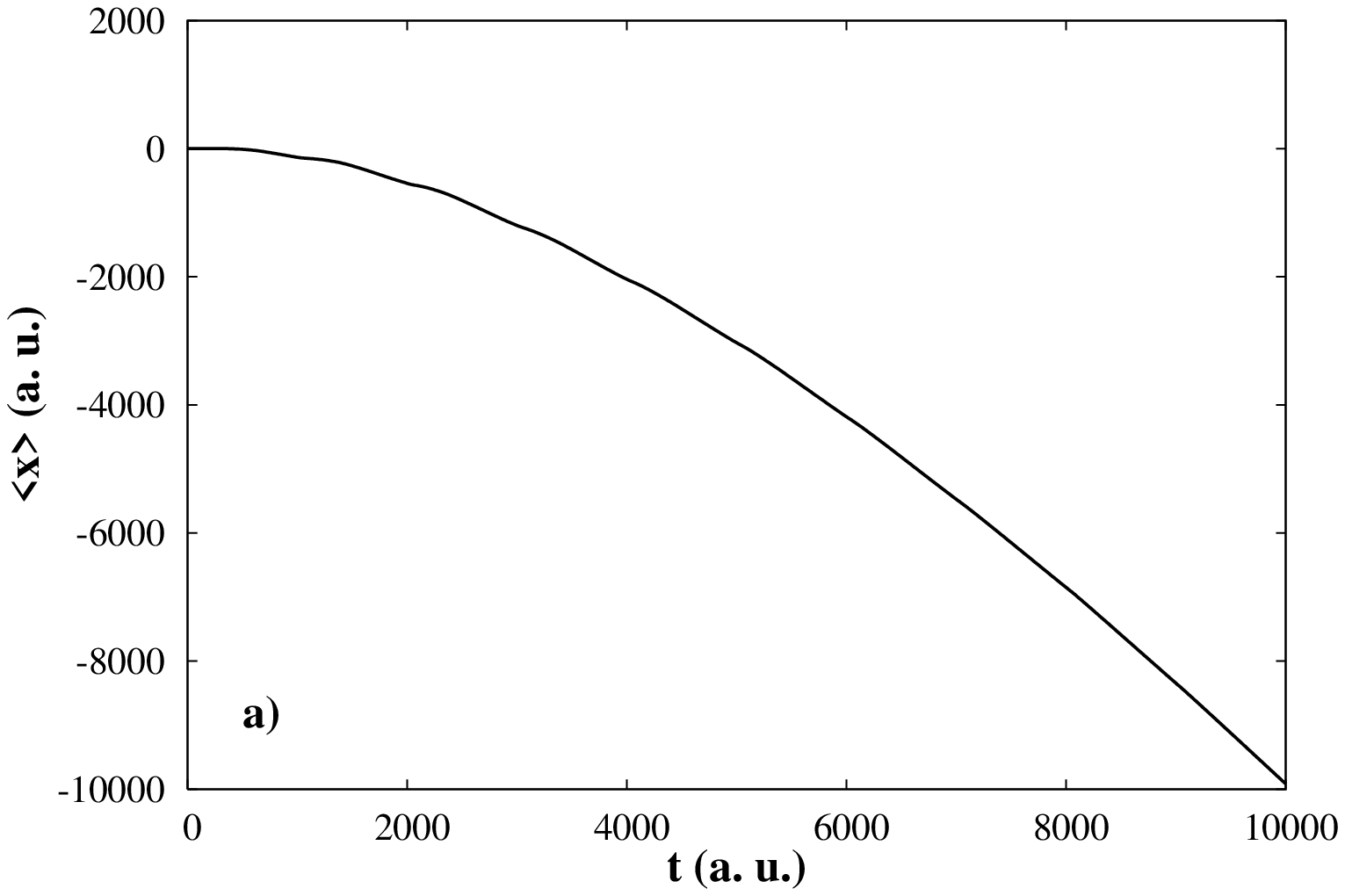}
\includegraphics[width=0.44\textwidth]{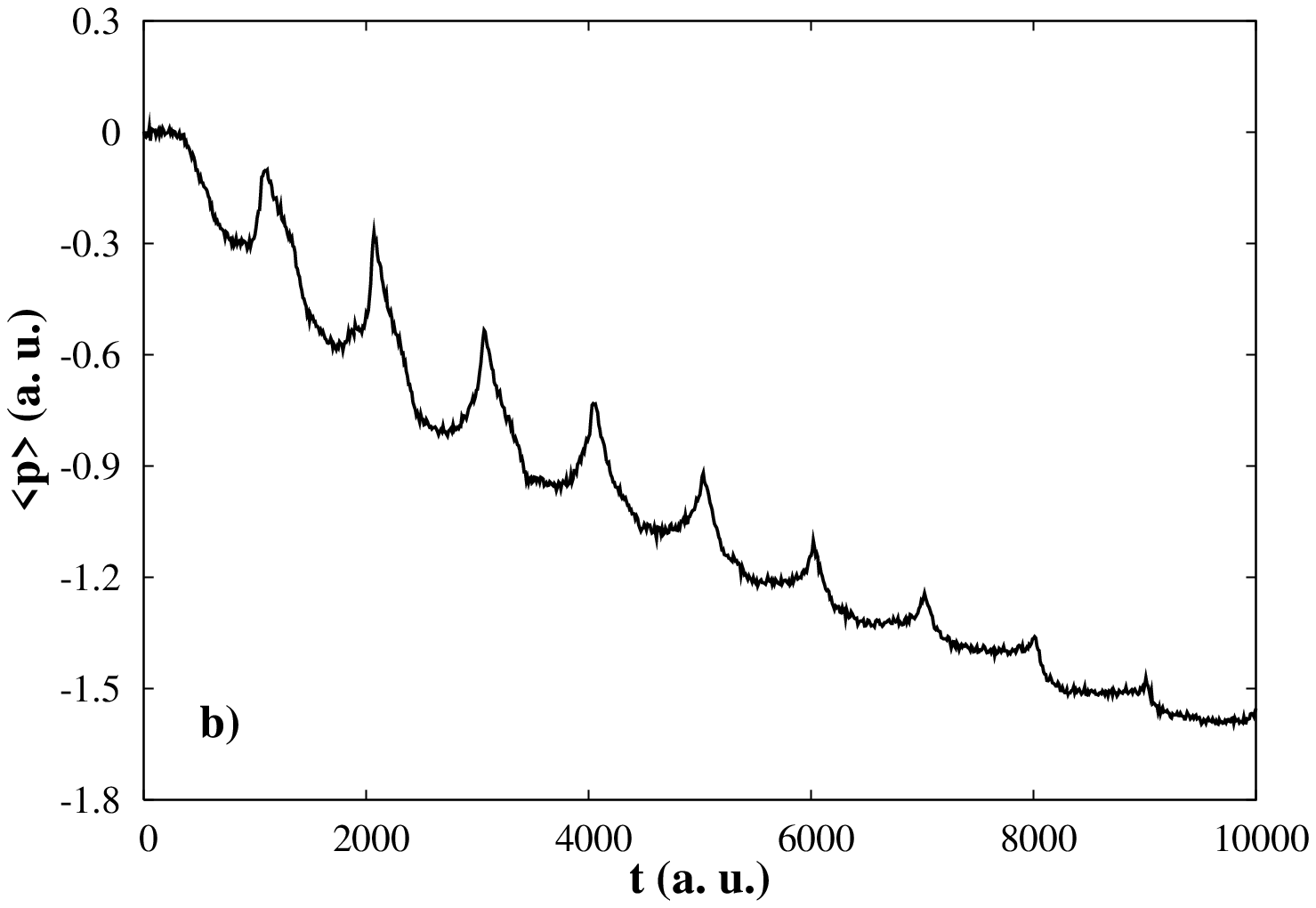}
\includegraphics[width=0.44\textwidth]{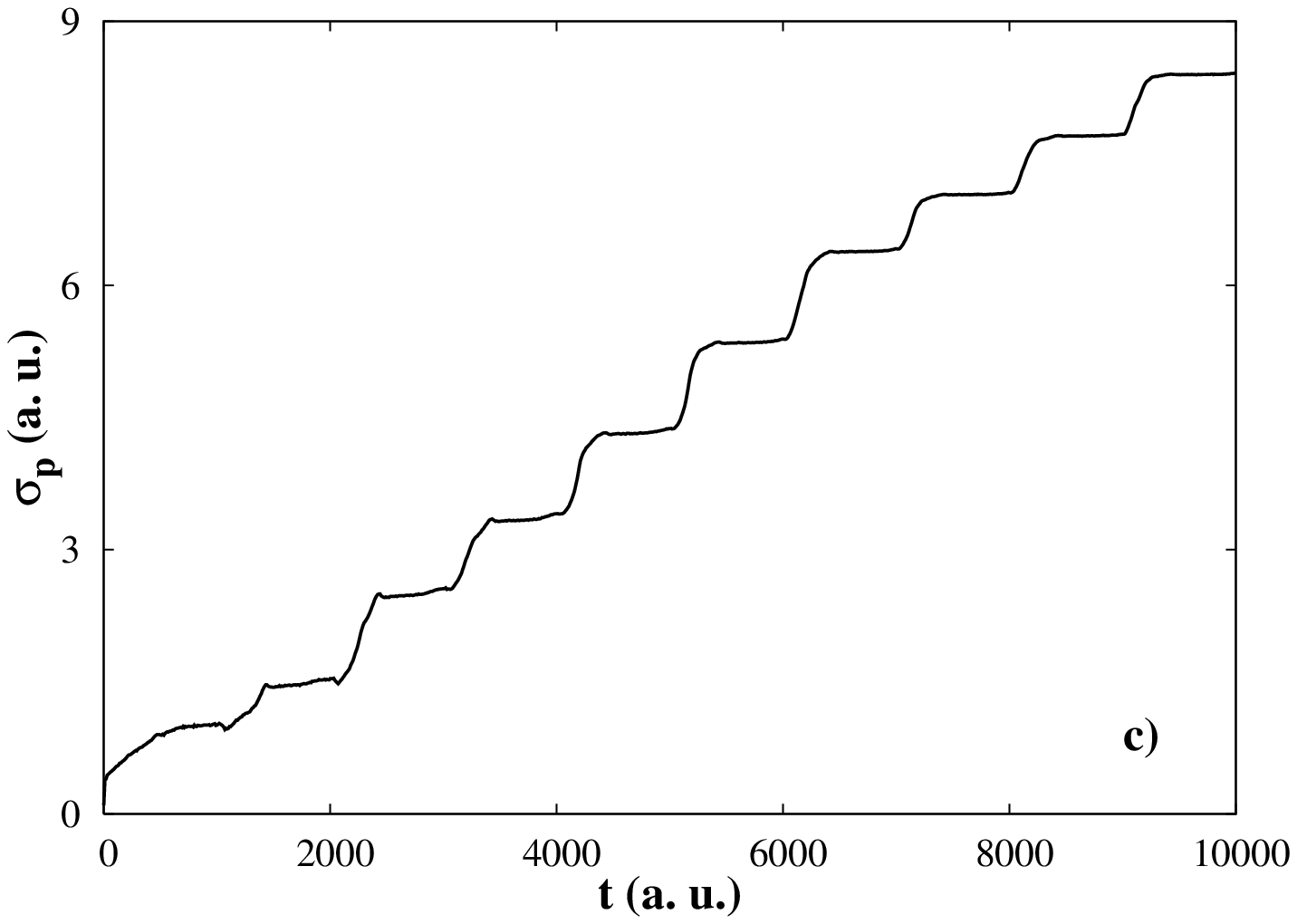}
\caption{(a) Mean coordinate, (b) mean momentum and
(c) momentum variance as functions of time.}
\label{fig2}
\end{center}
\end{figure}

It is natural to suggest that adiabatic variation of the resonant momentum
(\ref{res21}) leads to a gradual displacing
of areas of instability in phase space.
If the timescale of diffusive mixing inside the chaotic areas
is much smaller than the timescale of changing the resonant value
of momentum (\ref{res21}), then these areas play the role of dynamical traps
for particles, so-called stochastic layer traps (SLT) \cite{Rakhlin,Traps}.
Consequently displacement of a chaotic layer in energy space can be followed
by increasing or decreasing of mean energy of particles
belonging to it.
As it will be shown in this Letter,
a rather complicated situation occurs if wavenumber of the perturbation
varies according to the law (\ref{kt}).

\begin{figure}[!t]
\begin{center}
\includegraphics[width=0.4\textwidth]{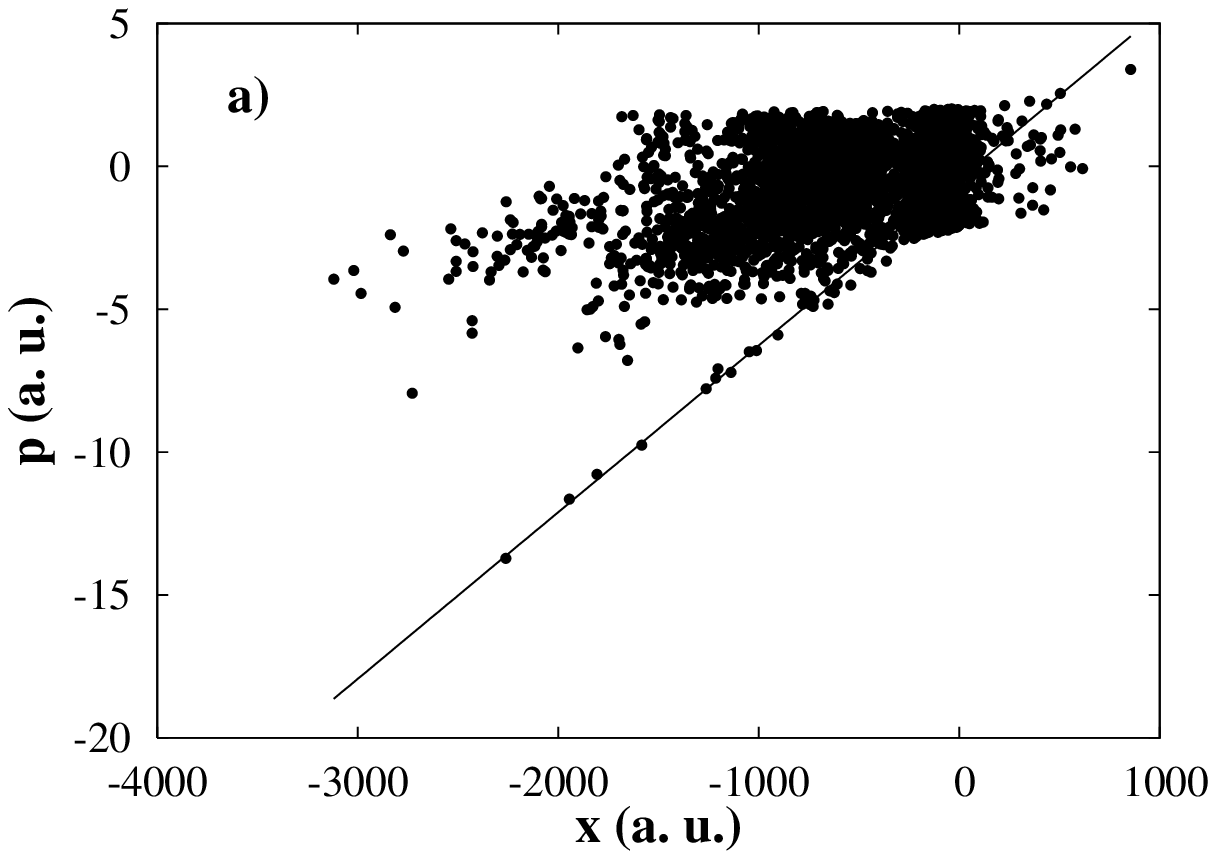}
\includegraphics[width=0.4\textwidth]{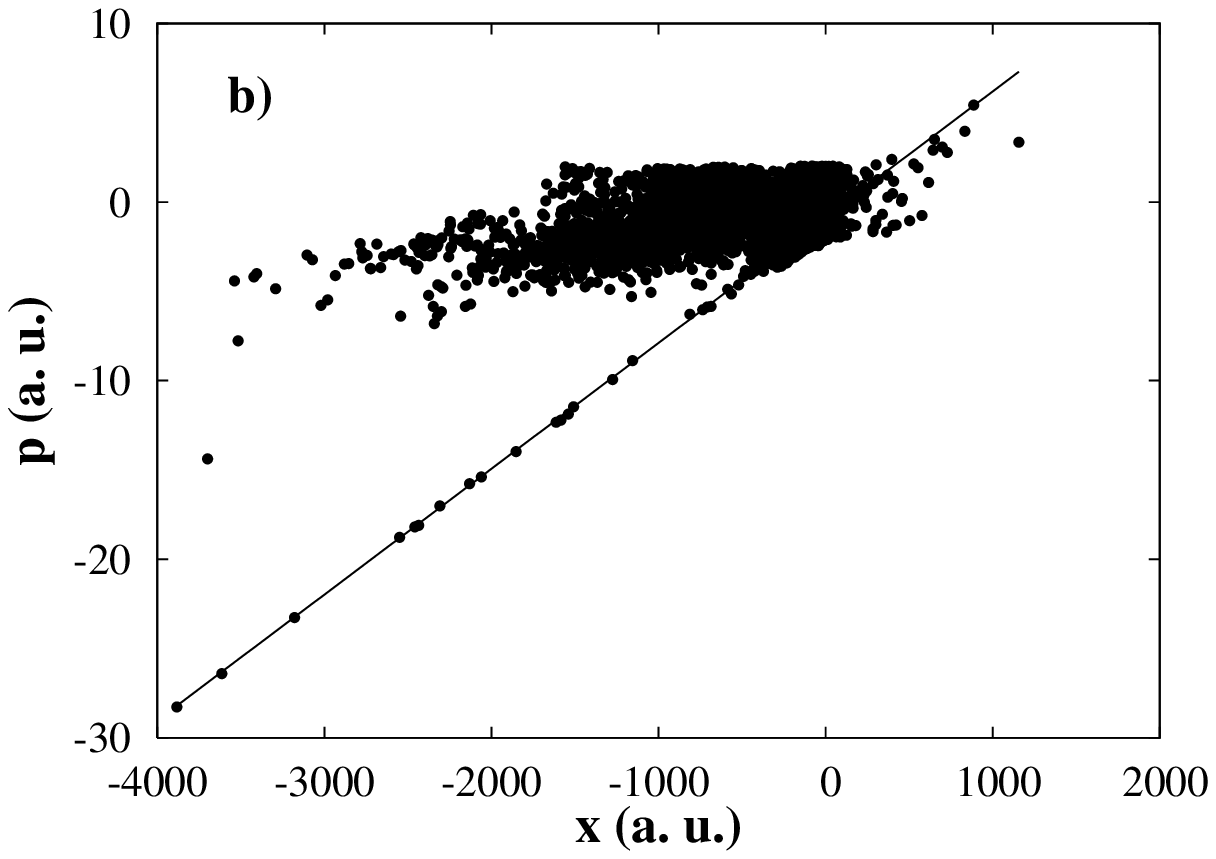}
\caption{Particle distribution in phase space at
(a) $t=1300$, (b) $t=1400$.
The resonant channel is marked by the line.
}
\label{fig3}
\end{center}
\end{figure}

Substituting (\ref{str}) into (\ref{f}),
we obtain the expression for the criterion (\ref{crit})
on the resonant line
\begin{equation}
\begin{aligned}
\biggl|\biggr.
\frac{a\Omega^2(-a\cos^2\Psi+\cos\Psi+2a)}{(1+a\cos\Psi)^2}x&
\\
+\sin x-\frac{2a\Omega\nu\sin\Psi}{k(1+a\cos\Psi)^2}
\biggl.\biggr|\le\varepsilon\tilde k&.
\end{aligned}
\label{crit2}
\end{equation}
This inequality holds if $\sin{x}\simeq 0$ and, subsequently,
$x\simeq \pi l$, where $l$ is integer. 
When skipped the term $\sim\sin{x}$, we can
rewrite the criterion (\ref{crit2}) as follows:
\begin{equation}
\begin{aligned}
g(x,\,\Psi)=&\biggl|\biggr.
\frac{a\Omega^2(-a\cos^2\Psi+\cos\Psi+2a)}{(1+a\cos\Psi)^2}x\\
&-\frac{2a\Omega\nu\sin\Psi}{k(1+a\cos\Psi)^2}
\biggl.\biggr|-\varepsilon\tilde k\le0.
\end{aligned}
\label{g}
\end{equation}
Figure \ref{fig1} represents the function $g(x,\,\Psi)$
with different fixed values of $x$.
According to this figure,
the criterion (\ref{crit2}) is satisfied with
$|x|<50000$ 
for the large intervals of $\Psi$, centered at $2\pi m$, where $m=0,1,2,...$.
This implies existence of those trajectories which, being passed the resonance
area at once,
will visit the resonant area repeatedly on the subsequent cycles of pendulum, till the 
slowly-varying phase $\Psi$ 
remains close to $2\pi m$. Such particles move along the lines described
by (\ref{str}), towards the point $x=0$ when $\sin{\Psi}<0$ and from it when $\sin{\Psi}>0$.
The latter ones are capable to
perform ballistic flights with increasing velocity.
\begin{figure}[!h]
\begin{center}
\includegraphics[width=0.4\textwidth]{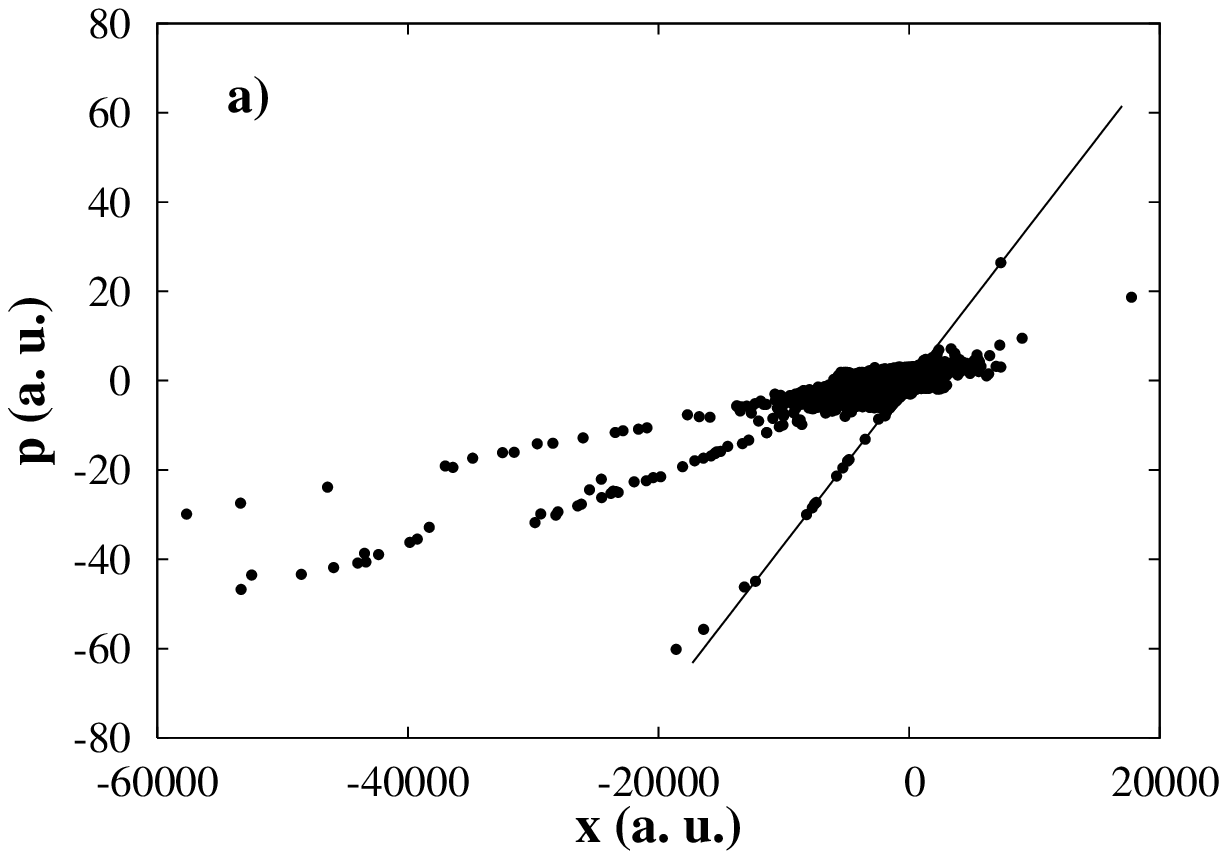}
\includegraphics[width=0.4\textwidth]{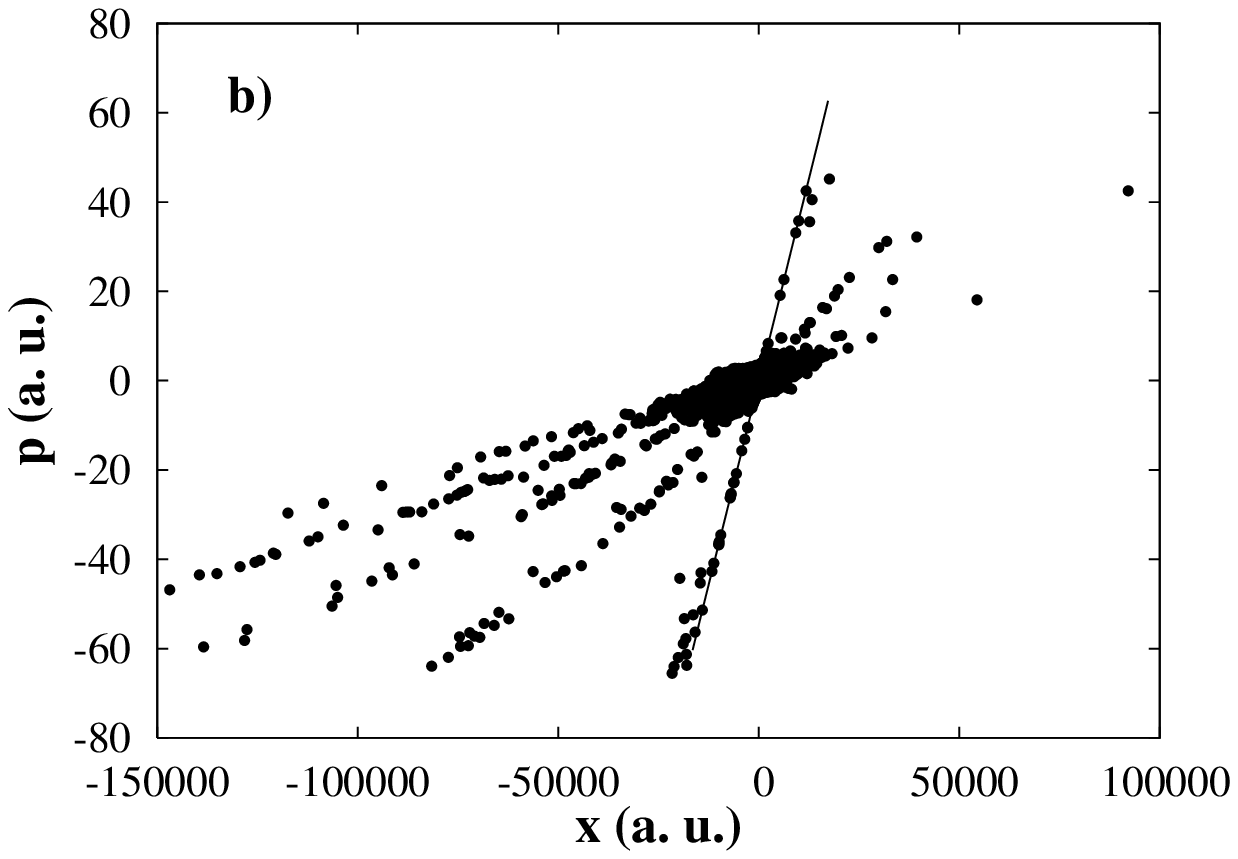}
\includegraphics[width=0.4\textwidth]{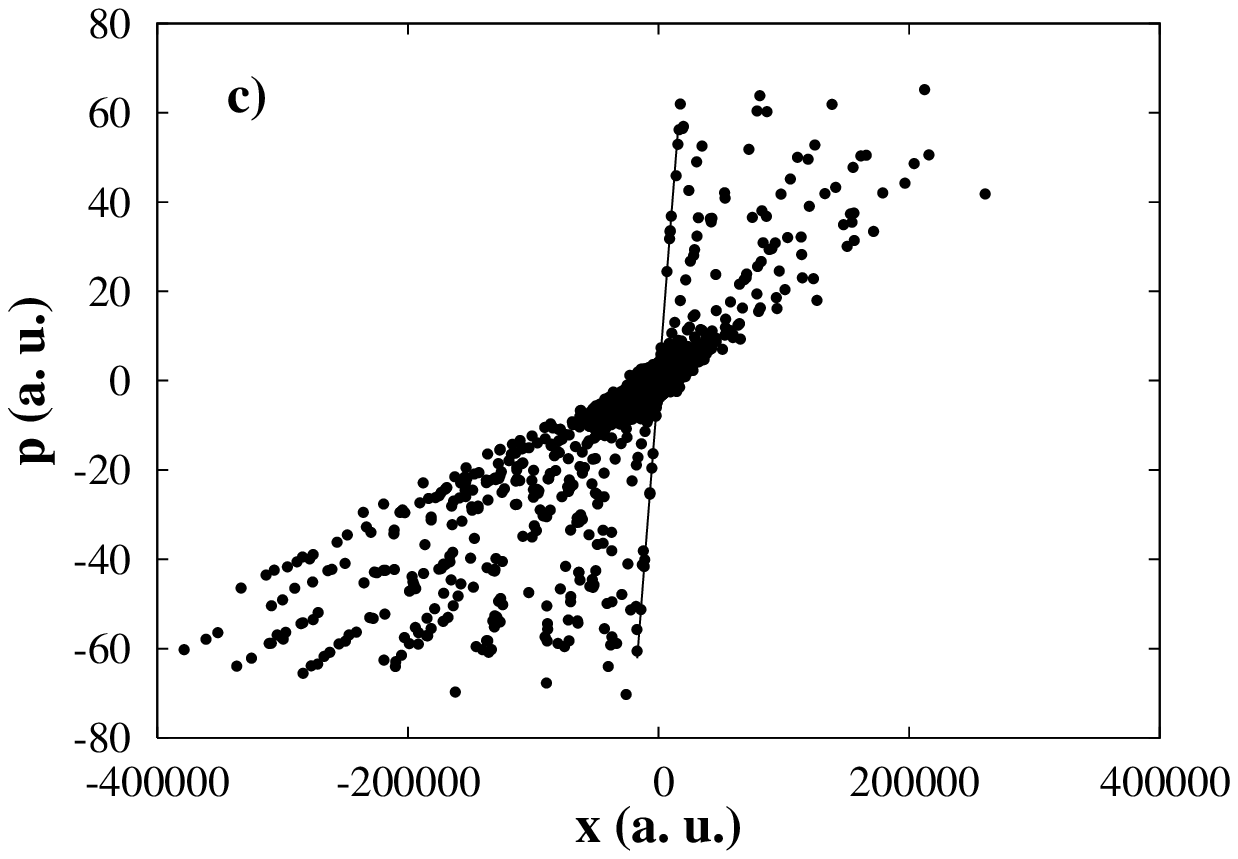}
\caption{The same as in Fig.~\ref{fig2} at (a) $t=3200$,
(b) $t=5200$, (c) $t=9200$.}
\label{fig4}
\end{center}
\end{figure}

Occurrence of such ballistic flights is confirmed by numerical
simulation.
We computed evolution
of the ensemble of particles,
initially distributed with gaussian probability density
\begin{equation}
\rho(x,\,p,\,t=0)=\frac{1}{2\pi\sigma_{0x}\sigma_{0p}}
\exp\left(-\frac{x^2}{\sigma_{0x}^2}-\frac{p^2}{\sigma_{0p}^2}\right),
\label{rho}
\end{equation}
where $\sigma_{0x}=\sigma_{0p}=0.1$.
The parameters of the perturbation we used are the following:
$\varepsilon=0.04$, $k_0=12$, $\nu=4$,
$\alpha=0.75$, $\Omega=2\pi/1000$.
Figure \ref{fig2} represents the temporal dependence of mean coordinate,
mean momentum and variance of momentum.
It is shown that there occurs a particle flux directed towards $t\to-\infty$.
The mean momentum grows nonmonotonically and abrupt accelerations
are alternating
with abrupt slowings down. Acceleration takes place when the slow phase
$\Psi$ is close
to $2\pi m$, that agrees with our analysis.
Each act of acceleration is followed by 
step-like increasing of momentum variance.
It should be emphasized that momentum variance
is much larger than mean momentum,
that indicates the presence of particles with very high velocities.
Figure~\ref{fig3} shows that accelerating particles form
jets along the resonant line (\ref{str}), which is cut
according to the criterion (\ref{g}).
The first significant jet becomes apparent at $t\simeq1300$.
Accelerating particles follow the resonant line until
$t\simeq1400$ and then leave the resonant zone.
It should be noted that strongly accelerated particles never
return into the resonant zone again.
Forming of later jets is demonstrated in Fig.~\ref{fig4},
where instantaneous particle distributions at $t=3200$,
$t=5200$ and $t=9200$ are presented.
Evolution of the particle cloud is also presented in the media files,
which are available at \cite{Clips}.

We can distinguish three stages of evolution of the particle ensemble.
At the first stage directed current is activating.
Until a particle is not far from $x=0$, the location of the resonant zone
is determined by the formula (\ref{res2}),
i.e. by the instantaneous value of the ratio $\nu/k(t)$.
It infers that the initial particle cloud is placed inside the chaotic layer
caused by resonance with $p_\text{res}=-\nu/\tilde k(t=0)=-4/21$. 
Adiabatic decreasing of $k$ displaces
this layer to the separatrix.
The time of diffusive mixing inside the chaotic layer
is much smaller than $2\pi/\Omega$, so that
the particle cloud follows the chaotic layer.
At $\Psi=\pi$ chaotic layer caused by resonance (\ref{res1}) merges into
the near-separatrix chaotic layer,
that leads to occurrence of ballistic particle current towards $x=-\infty$.

The second stage starts when the particle cloud becomes enough wide and some particles are
capable to fall into the resonant channels described by (\ref{str}).
This stage is characterized by fast growth of momentum variance
due to events of giant acceleration. 
Note that some particles turn around and then perform
ballistic flights in the opposite direction.
Nevertheless, number of particles
accelerating in the direction $x\to-\infty$
is much larger, therefore the turned particles
give negligible contribution into the resulting particle flux.
Since the resonant channels have finite length, one can call
the zone where they exist as the {\it accelerating zone}.

The third stage is not presented in the figures.
It starts when the particle cloud becomes very wide
and only negligible fraction
of a particle ensemble remains within the accelerating zone.
At this stage momentum variance achieves saturation
and stops increasing.

In conclusion, in this work we demonstrated the
effect of giant particle acceleration in the simple space-periodic
Hamiltonian
system subjected to a slowly-modulated external force.
The effect arises from the specific topology
of resonance (\ref{res1}) in phase space, which
permits capturing of a particle into the accelerating channel.

 
This work was supported by the projects of the President of the Russian Federation,
by the Program ``Mathematical Methods in 
Nonlinear Dynamics'' of the Prezidium of the Russian Academy of Sciences, 
and by the Program for Basic Research of the Far Eastern Division of 
the Russian Academy of Sciences. 
Authors are grateful to A.I.~Neishtadt, S.~Flach and S.V.~Prants
for helpful discussions during the course of this research.

\end{document}